\documentclass[conference]{IEEEtran}
\IEEEoverridecommandlockouts

\usepackage{cite}
\usepackage{amsmath,amssymb,amsfonts}
\usepackage{graphicx}
\usepackage{grffile}
\graphicspath{{images/}}
\usepackage{textcomp}
\usepackage{xcolor}
\usepackage{booktabs}
\usepackage{multirow}
\def\BibTeX{{\rm B\kern-.05em{\sc i\kern-.025em b}\kern-.08em
    T\kern-.1667em\lower.7ex\hbox{E}\kern-.125emX}}

\begin{document}

\title{Breaking the Pair: Evaluating dyadic interaction via speaker switching}


\author{
\IEEEauthorblockN{Nishchay Nilabh, Neeraj Kumar Sharma}
\IEEEauthorblockA{Mehta Family School of Data Science and Artificial Intelligence\\
Indian Institute of Technology Guwahati, 781039, India\\
\texttt{n.nishchay@alumni.iitg.ac.in, neerajs@iitg.ac.in}}
}

\maketitle

\begin{abstract}
Speakers in dialogue continuously adapt their communicative behavior across acoustic, lexical, and semantic dimensions, a phenomenon known as conversational entrainment. Modeling this process requires representations that capture the global structure of interaction, yet prior approaches fail to disentangle dyad-specific patterns from speaker-specific traits, limiting their ability to capture true conversational adaptation. We address this with the Dyadic Distance Matrix (DDM), which encodes all pairwise distances between the turns of two speakers over an entire conversation, capturing long-range cross-speaker dependencies. This raises a key question: does the DDM represent genuine interaction, or merely reflect individual speaker characteristics? We propose the speaker-switch test, a principled control in which one speaker’s turns are replaced with those from an unrelated speaker drawn from a different conversation. This preserves turn-level statistics while disrupting the original dyadic co-adaptation. The ability to distinguish real from switched DDMs thus directly evaluates whether the representation encodes interaction-specific structure. Across four embedding types and four classifiers (MLP, CNN, ResNet-50 and ViT) on the CANDOR corpus, real DDMs are consistently distinguishable from their switched counterparts, and this discriminability is robust to a fully speaker-disjoint data split. Comparisons with synthetic conversations created using LibriSpeech show higher discriminability in read speech, highlighting the role of prosodic variability in naturalistic conversations. Grad-CAM analysis further reveals distinct structural signatures driving classification. These results establish the speaker-switch test as a robust diagnostic for validating representations of dyadic conversational interaction.

\end{abstract}


\section{Introduction}
Speakers in dialogue continuously adapt their communicative behavior to their interlocutors across acoustic, lexical, and semantic dimensions~\cite{levitan2011, brennan1996, levitan2012}. This phenomenon, known as \emph{conversational entrainment}, is formalized under Communication Accommodation Theory (CAT)~\cite{giles1973,giles1991}, which predicts both convergence and divergence depending on communicative goals. Capturing entrainment computationally therefore requires representations that move beyond local, adjacent-turn similarity to model the evolving global structure of an entire conversation. However, most prior computational approaches rely on local turn-level metrics~\cite{nasir2022, lahiri2023}, which conflate two distinct sources of structure: (i)~\emph{conversation-specific} patterns arising from the joint behavior of a dyad, and (ii)~\emph{speaker-specific} characteristics intrinsic to each participant. As a result, models may rely on speaker identity rather than true dyadic co-adaptation, limiting their validity as measures of conversational entrainment.

To address this limitation, we use the Dyadic Distance Matrix (DDM), a representation that encodes all pairwise cosine distances between the turn embeddings of two speakers across an entire conversation. By capturing long-range cross-speaker dependencies and global temporal structure, the DDM provides a compact representation of interaction dynamics. This naturally raises a fundamental question: does the DDM encode genuine \emph{interaction}, or does it primarily reflect the statistical properties of individual speakers?

\begin{figure*}[!t]
\centering
\includegraphics[width=\textwidth]{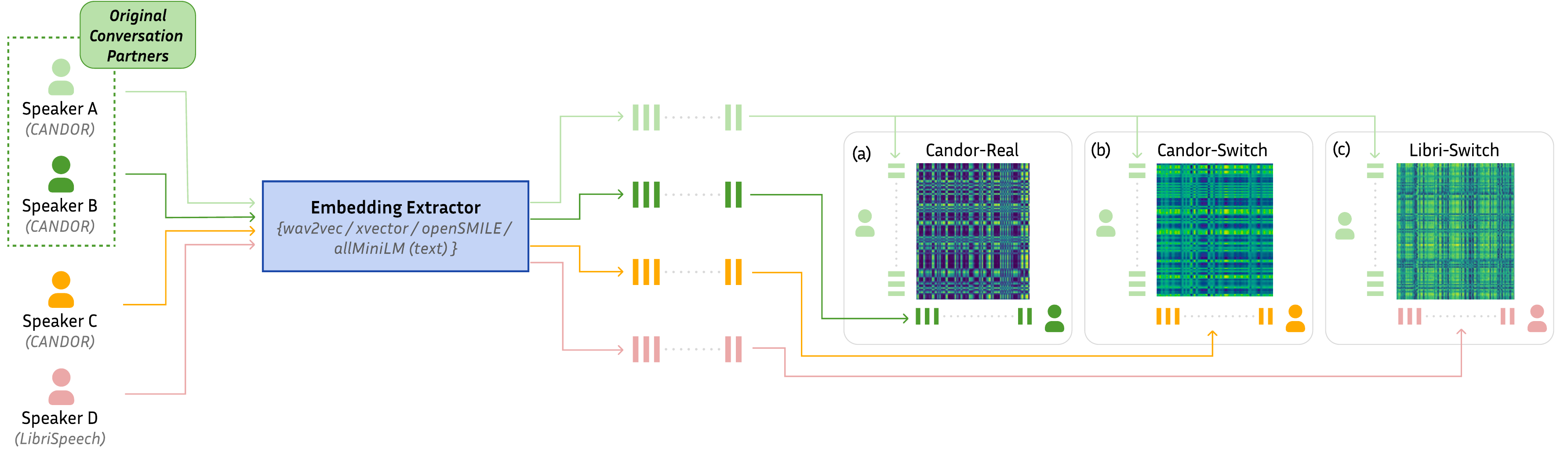}
\caption{Dyadic Distance Matrix (DDM) generation pipeline. Turn-level embeddings ($a_i, b_j$) are extracted using acoustic or semantic models to compute the global cross-speaker distance matrix ($M$). The schematic illustrates the formulation of the three experimental pairings: (a) genuine interaction between original conversation partners (CANDOR-Real), (b) within-corpus control using an unrelated interlocutor (CANDOR-Switch), and (c) the cross-corpus read-speech (Libri-Switch) using Speaker D's speech from the LibriSpeech dataset~\cite{panayotov2015}.}
\label{fig:main}
\end{figure*}

We answer this question through the \emph{speaker-switch test} (Fig.~\ref{fig:main}), a principled control designed to disrupt dyadic interaction while preserving speaker-level properties. Given a conversation between speakers $A$ and $B$, we construct a synthetic counterpart by replacing $B$’s turn sequence with turns from an unrelated speaker drawn from a different conversation. This produces a controlled contrast (e.g., CANDOR-Real vs.\ CANDOR-Switch) that maintains individual feature distributions and turn structure, while eliminating genuine co-adaptation between the original pair. A model that reliably distinguishes real from switched DDMs therefore provides direct evidence that the representation encodes interaction-specific structure. Our contributions are as follows:
\begin{enumerate}
\item We introduce the speaker-switch test, a principled evaluation framework for distinguishing interaction-specific structure from speaker-specific effects in conversational representations.
\item We conduct a systematic evaluation across four embedding types spanning acoustic and semantic layers, comparing ResNet-50, CNN, MLP, and ViT classifiers.
\item We perform a cross-corpus analysis using the LibriSpeech read-speech dataset~\cite{panayotov2015} to examine how prosodic variability in naturalistic speech affects acoustic discriminability.
\item We provide a Grad-CAM-based~\cite{selvaraju2017} interpretability analysis identifying which structural regions of the DDM carry discriminative signal.
\end{enumerate}

\section{Background}

\subsection{Dyadic Distance Matrix}
\noindent We represent a two-speaker conversation using the Dyadic Distance Matrix (DDM), which captures cross-speaker similarity across all turns. Let ${\mathbf{a}_i}$ and ${\mathbf{b}_j}$ denote the embedding sequences corresponding to the turns of speakers $A$ and $B$, respectively. The DDM is defined as a matrix $M \in \mathbb{R}^{|\mathbf{a}| \times |\mathbf{b}|}$, where each entry $(i,j)$ encodes the cosine distance between turns $i$ and $j$:

\begin{equation}
M_{i,j} = 1 - \frac{\mathbf{a}_i^\top \mathbf{b}_j}
{\|\mathbf{a}_i\|_2\,\|\mathbf{b}_j\|_2}.
\label{eq:ddm}
\end{equation}

Each entry of $M$ reflects the distance (dissimilarity) between a pair of turns across speakers, providing a global view of interaction over the entire conversation. Unlike adjacent-turn metrics~\cite{nasir2022, lahiri2023}, which focus on local dependencies, the DDM encodes all cross-speaker turn combinations, enabling the analysis of long-range interactional structure.

Following Communication Accommodation Theory (CAT), we compute DDMs across three complementary representational layers:
(i)~an \emph{acoustic layer} (wav2vec~2.0, openSMILE, x-vector), capturing prosodic, paralinguistic and speaker-specific cues; and
(ii)~a \emph{semantic layer} (all-MiniLM), capturing lexical and topical alignment between speakers.

\section{Methodology}

\subsection{Datasets}
\noindent We use the CANDOR corpus~\cite{reece2023}, comprising 1,656 spontaneous dyadic conversations (approximately 850 hours) from over 1,400 speakers recorded via video-calling platforms. Each session lasts approximately 31 minutes and includes word-level timestamps, speaker diarization, and transcripts. We apply Cliffhanger sentence-level segmentation to obtain semantically coherent speaker turns, and exclude sessions with fewer than 20 turns per speaker.

To assess the role of prosodic variability, we conduct a complementary cross-corpus experiment using LibriSpeech~\cite{panayotov2015}. As a large-scale read-speech corpus, LibriSpeech provides a controlled setting with relatively uniform prosody, in contrast to the highly variable naturalistic speech in CANDOR.

\subsection{Task Formulation}
\noindent We cast the speaker-switch test as a binary classification task over DDMs. Given a Dyadic Distance Matrix $M$, the objective is to predict whether it is derived from a \emph{real} conversation (CANDOR-Real) or a \emph{speaker-switched} counterpart (CANDOR-Switch). 

Formally, the classifier learns a mapping $f: M \rightarrow \{0,1\}$, where the label indicates the presence or absence of genuine dyadic interaction. Successful discrimination implies that the DDM encodes interaction-specific structure beyond speaker-level statistics.

\subsection{Speaker-Switch DDM Construction}
\noindent Given an original conversation $\mathcal{C}_1 = (A_1, B_1)$ and an unrelated conversation $\mathcal{C}_2 = (A_2, B_2)$, the switched DDM for $\mathcal{C}_1$ is constructed by replacing $B_1$ with $B_2$, while retaining $A_1$'s turn sequence. Conversation pairs are matched by approximate turn count to preserve matrix dimensions.

\begin{equation}
    M^{\mathrm{sw}}_{i,j} = 1 - \frac{{\mathbf{a}^{(1)}_i}^\top \mathbf{b}^{(2)}_j}
                        {\|\mathbf{a}^{(1)}_i\|_2\,\|\mathbf{b}^{(2)}_j\|_2}.
\end{equation}

This construction preserves: (i) each speaker's embedding distribution; (ii) within-conversation temporal structure; and (iii) global pairwise distance statistics, while eliminating (iv) genuine dyadic co-adaptation between $A_1$ and $B_1$.

\subsection{Feature Extraction}
\noindent We evaluate four embedding types corresponding to the three CAT representational layers:

\begin{itemize}
\item \textbf{wav2vec~2.0}~\cite{baevski2020}: 768-dimensional representations obtained via mean-pooling of the final transformer layer; the base pre-trained model is used to retain prosodic information.
\item \textbf{x-vector}~\cite{desplanques2020}: 512-dimensional speaker embeddings extracted from a pre-trained ECAPA-TDNN model via \texttt{pyannote.audio}~\cite{bredin2023pyannote}.
\item \textbf{openSMILE (eGeMAPS)}~\cite{eyben2010,eyben2016}: 88-dimensional acoustic descriptors capturing pitch, energy, spectral, and cepstral features.
\item \textbf{all-MiniLM}~\cite{reimers2019}: 384-dimensional sentence embeddings encoding contextual semantic information.
\end{itemize}

Per-speaker $z$-normalization is applied to wav2vec~2.0, x-vector, and openSMILE embeddings to remove static speaker-level biases prior to DDM computation.

\subsection{Classifiers}
\noindent We evaluate four classifier architectures. Because each DDM is a two-dimensional image-like array whose discriminative cues are spatially distributed (Sec.~\ref{sec:gradcam}), our primary model is a ResNet-50 backbone~\cite{he2016} (in PyTorch~\cite{paszke2019}), whose hierarchical residual features capture multi-scale spatial structure; its final fully connected layer is replaced by a dropout layer ($p=0.3$) and a sigmoid output. A three-layer CNN and a multilayer perceptron (MLP) serve as shallow baselines, and a ViT-B/16~\cite{dosovitskiy2021} (ImageNet-pretrained, $16\times16$ patches) provides a globally aware, attention-based comparison.

All models use the Adam optimizer (learning rate $1\times10^{-4}$, batch size 32) with early stopping on validation loss. The data are split 70/15/15 (train/validation/test) stratified by conversation; DDMs are resized via bilinear interpolation to $64\times64$ ($224\times224$ for ViT) and standardized to unit variance. We report accuracy, macro-F1, and Equal Error Rate (EER), the threshold-independent operating point at which false-acceptance and false-rejection rates coincide. To rule out speaker-identity leakage, we additionally re-evaluate under a fully speaker-disjoint split (no speaker shared across train/validation/test): discriminability is essentially unchanged, with accuracy, macro-F1, and EER all remaining within a small margin of the standard-split values across all modalities, and a probe stratifying test pairs by whether their speakers appear in training performs near-identically on seen and unseen speakers. This confirms the classifiers capture interaction structure rather than speaker identity.

\subsection{Cross-Corpus Evaluation}
\noindent To isolate the effect of prosodic variability, we construct cross-corpus switched DDMs by replacing one speaker in a CANDOR conversation with a randomly sampled speaker from LibriSpeech. Specifically, given a CANDOR conversation $\mathcal{C}_1 = (A_1, B_1)$, we retain $A_1$’s turn sequence and replace $B_1$ with speech segments from an unrelated LibriSpeech speaker, matched by approximate turn count.

This results in a synthetic interaction between naturalistic conversational speech and read speech, preserving the structure of one side of the dialogue while introducing a controlled mismatch in speaking style. Applying the same classification pipeline allows us to assess how reduced prosodic variability in read speech affects discriminability relative to fully naturalistic dyadic interactions.

\subsection{Grad-CAM Interpretability}
\noindent We apply Gradient-weighted Class Activation Mapping (Grad-CAM)~\cite{selvaraju2017} to the trained ResNet-50 to identify discriminative regions within the DDM. For each correctly classified test sample, Grad-CAM heatmaps are computed and averaged across samples within each condition and embedding type. These aggregated maps reveal the structural loci of interaction-specific information captured by the model.

\section{Results and Discussion}

\subsection{Classification: Real vs.\ Switch}

\noindent Table~\ref{tab:classification} presents classification results across all
modalities, model architectures, and corpora. On CANDOR, the semantic
embedding (all-MiniLM) achieves perfect discrimination with ResNet-50
(Acc=1.000, EER=0.000), and remains highly discriminable even with the
shallow CNN (Acc=0.952) and MLP (Acc=0.857). Semantic alignment is thus
the most easily detected signal, robust to classifier choice. However, as
we discuss below, this discriminability is partly attributable to topical
divergence between unrelated conversations rather than to fine-grained
co-adaptation alone.

\begin{table*}[t]
\caption{Real vs.\ Switch classification across models and modalities. Best per metric/modality/corpus in \textbf{bold}.}
\label{tab:classification}
\begin{center}
\small
\renewcommand{\arraystretch}{1.15}
\begin{tabular*}{\textwidth}{@{\extracolsep{\fill}} l l cccc cccc @{}}
\toprule
& & \multicolumn{4}{c}{\textbf{CANDOR-Switch}}
  & \multicolumn{4}{c}{\textbf{Libri-Switch}} \\
\cmidrule(lr){3-6}\cmidrule(lr){7-10}
\textbf{Modality} & \textbf{Metric}
 & \textbf{MLP} & \textbf{CNN} & \textbf{ResNet-50} & \textbf{ViT}
 & \textbf{MLP} & \textbf{CNN} & \textbf{ResNet-50} & \textbf{ViT} \\
\midrule
\multirow{3}{*}{wav2vec 2.0}
 & Acc   & 0.584 & 0.552 & 0.620 & \textbf{0.647} & \textbf{1.000} & 0.823 & 0.904 & 0.918 \\
 & F1    & 0.559 & 0.433 & 0.649 & \textbf{0.651} & \textbf{1.000} & 0.820 & 0.904 & 0.919 \\
 & EER   & 0.402 & 0.474 & 0.394 & \textbf{0.353} & \textbf{0.000} & 0.169 & 0.092 & 0.080 \\
\midrule
\multirow{3}{*}{x-vector}
 & Acc   & 0.500 & 0.597 & \textbf{0.682} & 0.658 & \textbf{1.000} & \textbf{1.000} & \textbf{1.000} & \textbf{1.000} \\
 & F1    & 0.000 & 0.572 & \textbf{0.687} & 0.663 & \textbf{1.000} & \textbf{1.000} & \textbf{1.000} & \textbf{1.000} \\
 & EER   & 0.485 & 0.388 & \textbf{0.315} & 0.339 & \textbf{0.000} & \textbf{0.000} & \textbf{0.000} & \textbf{0.000} \\
\midrule
\multirow{3}{*}{openSMILE}
 & Acc   & 0.467 & 0.575 & \textbf{0.692} & 0.554 & \textbf{0.975} & 0.842 & 0.908 & 0.942 \\
 & F1    & 0.458 & 0.557 & \textbf{0.722} & 0.550 & \textbf{0.974} & 0.832 & 0.904 & 0.940 \\
 & EER   & 0.533 & 0.450 & \textbf{0.350} & 0.467 & \textbf{0.000} & 0.150 & 0.083 & 0.067 \\
\midrule
\multirow{3}{*}{all-MiniLM}
 & Acc   & 0.857 & 0.952 & \textbf{1.000} & 0.744 & \textbf{1.000} & 0.994 & 0.998 & 0.998 \\
 & F1    & 0.858 & 0.950 & \textbf{1.000} & 0.739 & \textbf{1.000} & 0.994 & 0.998 & 0.998 \\
 & EER   & 0.149 & 0.064 & \textbf{0.000} & 0.211 & \textbf{0.000} & 0.004 & 0.004 & 0.004 \\
\bottomrule
\end{tabular*}
\end{center}
\end{table*}

\begin{figure*}[!t]
    \centerline{\includegraphics[width=\textwidth]{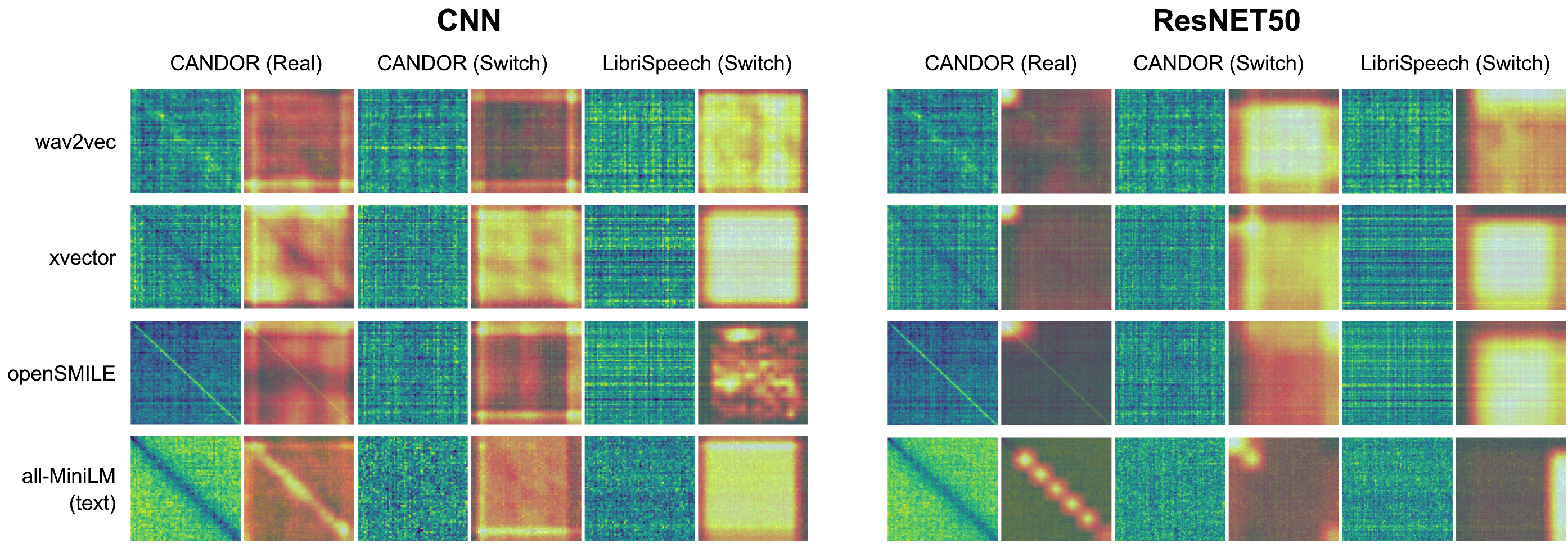}}
    \caption{Mean Grad-CAM heatmaps for ResNet-50 and CNN models across all embedding types (columns: wav2vec~2.0, x-vector, openSMILE, all-MiniLM). Rows show averaged input DDMs and Grad-CAM activations for real and switch conditions, where warmer colors indicate higher activation.}

    \label{fig:gradcam}
\end{figure*}

For acoustic embeddings on CANDOR, ResNet-50 consistently
outperforms the shallower architectures. x-vector ResNet-50 achieves
Acc=0.682 versus CNN at 0.597, while the MLP collapses to majority-class
prediction (Acc=0.500, F1=0.000, i.e.\ it assigns all samples to a single
class), unable to learn the interaction patterns encoded in the structural DDM. A similar collapse occurs for openSMILE MLP (Acc=0.467, below chance), whereas ResNet-50 attains 0.692. Notably, the ViT, despite its global self-attention, does not surpass ResNet-50 on CANDOR (x-vector 0.658 vs.\ 0.682; openSMILE 0.554 vs.\ 0.692), exceeding it only on wav2vec~2.0 (0.647 vs.\ 0.620). This indicates that hierarchical, locally aggregated spatial features, rather than global attention alone, best capture the distributed interaction structure, motivating ResNet-50 as our primary model.

That all four embedding types exceed chance on CANDOR with at least one model provides validation: DDMs encode genuine interaction-specific structure that cannot be recovered by combining arbitrary speakers. Semantic representations are the most discriminable, but this result is partly trivial: unrelated conversations differ in topic, so a switched semantic DDM loses topical continuity by construction. The near-perfect all-MiniLM score should therefore be read as an upper bound rather than as evidence of fine-grained semantic co-adaptation; the more informative evidence of interaction-specific structure comes from the acoustic embeddings, where topic is not a confounding cue. A stronger semantic control that pairs topic-matched conversations before switching is left to future work.

\subsection{Cross-Corpus Results: The Role of Prosodic Variability}

\noindent The contrast between CANDOR and LibriSpeech results is the most striking
finding in Table~\ref{tab:classification}. For x-vector embeddings, all
four classifiers achieve perfect accuracy on LibriSpeech (Acc=1.000,
EER=0.000), including the MLP that collapses on CANDOR. LibriSpeech
consists of read speech where intonation is controlled and within-speaker
variability is low. In this setting, x-vector embeddings robustly encode
stable, distinct speaker signatures, making cross-speaker pairing trivially detectable regardless of classifier capacity.

On CANDOR, speakers exhibit rich intonation variation, voice modulation,
and prosodic accommodation. Post-$z$-normalization, these residual patterns of within-speaker variability and cross-speaker adaptation reduce the discriminability of switched DDMs, and only ResNet-50 retains meaningful performance. The acoustic embedding results similarly follow this pattern: wav2vec~2.0 and openSMILE both show markedly higher accuracy on LibriSpeech than on CANDOR across all models. The LibriSpeech result thus serves as an important sanity check: the
difficulty of distinguishing real from switched acoustic DDMs on CANDOR
is not a failure of the representation, but rather a reflection of the
genuine complexity of conversational acoustic adaptation.

Notably, the MLP achieves Acc=1.000 on wav2vec~2.0 LibriSpeech despite
underperforming ResNet-50 on CANDOR. This inversion suggests that read speech embeddings contain a simple low-dimensional discriminative feature exploitable by a linear-style classifier, while conversational speech requires multi-scale spatial reasoning to separate real from switched patterns.

\subsection{Grad-CAM Analysis}
\label{sec:gradcam}

\noindent Fig.~\ref{fig:gradcam} shows mean Grad-CAM activations for ResNet-50 and CNN, aggregated over correctly
classified test samples under real and switch conditions.

\textbf{Semantic modality:}
For all-MiniLM, ResNet-50 Grad-CAM maps under the real condition exhibit
strong activation concentrated along and near the DDM diagonal. This
diagonal structure reflects temporal proximity: turns occurring close in
time are semantically similar, consistent with topic progression and shared reference in natural conversation. Under the switch condition, the diagonal activation disappears, replaced by diffuse or off-diagonal patterns, confirming that the diagonal is a signature of genuine conversational content alignment rather than any individual speaker property.

\textbf{Acoustic modality:}
For acoustic embeddings, ResNet-50 activates more globally rather
than concentrating on the diagonal. This is consistent with the literature documenting acoustic accommodation operating over longer time
scales~\cite{levitan2011}: prosodic adaptation is distributed across the
full conversation timeline rather than localized to adjacent turns. Notably, Grad-CAM visualizations for ResNET-50 real CANDOR DDMs display a localized concentration in the top-left quadrant; while this pattern warrants further investigation, the overall activation remains broadly distributed across the matrix.

\textbf{ResNet-50 vs.\ CNN:}
Within Fig.~\ref{fig:gradcam}, the CNN exhibits broader and less spatially structured activations than ResNet-50, particularly for acoustic modality. This explains the performance gap on CANDOR: the hierarchical feature extraction of ResNet-50 enables detection of complex, spatially distributed interaction signatures, whereas the three-layer CNN lacks sufficient depth.

\section{Conclusion}

We have presented a systematic speaker-switch validation of Dyadic Distance Matrices as representations of conversational interaction. Across four embedding types on CANDOR, switched DDMs are consistently distinguishable from real DDMs, validating that the DDM encodes genuine dyad-specific interaction structure beyond individual speaker characteristics. The LibriSpeech cross-corpus experiment shows that the lower discriminability of acoustic DDMs on CANDOR reflects the genuine prosodic variability of naturalistic conversation rather than a weakness of the representation. Grad-CAM analysis shows the structural signatures in the DDMs that drive classification: diagonal temporal alignment for semantic features and distributed global structure for acoustic features. These findings provide an interpretable foundation for using DDMs as a representation of conversational dynamics. Future work will integrate these validated DDM representations into predictive architectures. By isolating co-adaptation from speaker-specific traits, DDMs may enable applications such as outcome prediction, mutual understanding assessment, and more responsive dialogue systems.

\section*{Acknowledgment}

The authors thank Srikanth Raj Chetupalli and Shreyas Ramoji for discussions on this work.

\bibliographystyle{IEEEtran}
\bibliography{IEEEabrv, references}

\end{document}